\documentstyle[12pt]{article} 
\begin{document}

\title{Dynamics of Immobilized Flagella}

\author{D. Fry$^1$\footnote{\uppercase{E}-mail fryd@nsula.edu}, N. Hutchings$^2$\footnote{\uppercase{E}-mail hutchingsn@nsula.edu}, and A. Ludu$^1$\footnote{\uppercase{E}-mail ludua@nsula.edu}
Northwestern State University \\
$^1$Department of Chemistry and Physics,  $^2$Department of Biology, \\
Natchitoches, LA 71497, USA\\ 
E-mail: ludua@nsula.edu \\}

\date{\today}
\maketitle

\abstract{Although the auger-like 'swimming' motility of the African trypanosome was described upon its discovery over one hundred years ago, the precise biomechanical and biophysical properties of trypanosome flagellar motion has not been elucidated.  In this study, we describe five different modes of flagellar beat/wave patterns in African trypanosomes by microscopically examining the flagellar movements of chemically tethered cells.  The dynamic nature of the different beat/wave patterns suggests that flagellar motion in Trypanosoma brucei is a complex mixture of oscillating waves, rigid bends, helical twists and non-linear waves.   Interestingly, we have observed soliton-like depression waves along the flagellar membrane, suggesting a nonlinear mechanism for the dynamics of this system. The physical model is inspired by the 2-dimensional elastic dynamics of a beam, and by taking into account uniform distribution of molecular motors torque and nonlinear terms in the curvature.}

\section{Introduction}
Understanding the biomechanics and biophysics of cellular motility is of vital importance to advance our knowledge of how the cell's use their unique cytoarchitecture to sense, respond, and adapt to their surroundings. However, without a strong foundational understanding of how the cytoarchitecture facilitates the various states of cell motility, we can only superficially appreciate how motile behavior depends on intercellular and extracellular stimuli.  A large volume of work currently describes cellular architecture in both motile and non-motile cells.  However, a high-resolution understanding of dynamic motility requires us to extend the various cell-biological microscopy techniques that are used to examine nonmotile and 'fixed' cells to facilitate investigating real-time behavior of highly motile cells.  Through technological advancements in cellular immobilization (tethering), careful biological observations, and physical and mathematical modeling, we can apply our understanding of cellular architecture to elucidate the biomechanical properties of motile cells.  Research in cell motility has implications on our understanding biological complexity, cell behavior, drug design and delivery, infection and disease modeling, and may provide critical insight into engineering potential novel propulsion solutions for future.
	In addition to benefits for the biological sciences, the disciplines of mathematics, physics, organic chemistry, and analytical chemistry also share the rewards from studies in dynamic cell motility.  For example, understanding how cells use their unique shape and surface chemistry to interact with other cells and surfaces is a question best solved by investigative analytical chemistry.  Cell motility is an ideal model system in which to observe and mathematically model complex waves and to physically elucidate how a cell's environment affects its biomechanical properties.  By conducting interdisciplinary research on individual living cells, we are seeking to complete the following experimental objectives: 1) successfully tether the insect stage of the monoflagellated protozoan parasite, Trypanosoma brucei, to chemically modified glass to enable high-resolution visualization of trypanosome flagellar motion; 2) apply molecular and cell biological tool to describe the different types of flagellar motion that can be observed in procyclic form Trypanosoma brucei; and 3) use recent advancements in linear and non-linear waves to mathematically model the nature of the various flagellar beat patterns observed by quantifying the movements of the flagella and simulating flagellar beating in-silica.
There are currently several established and several emerging eukaryotic model systems available in which to examine the biophysical/biochemical properties of flagellar motion.  There are additional thousands of different species of flagellated cells that have been classified, each with specific nuances regarding the structure and function of the eukaryotic flagellum.  At this early stage in the establishment of core principles regarding cell motility, it is imperative to consider the work from each model system and try to bridge the principles and observations between the various systems.  Conversely, it is important to utilize the technologies and unique biology of each model system to conduct creative and intuitive studies within that particular model system regarding flagellated motility.
The biflagellated green algae, Chlamydomonas rheinhardti, is an established system for using classical genetics to studies the genes that encode flagellar constituents and their respective protein's function within the flagellum.  There are over 50 different motility mutant cell lines currently described by their structural and/or functional role in motility ~\cite{nate1}.  The Chlamydomonas flagella are comprised of a membrane-covered axoneme that is attached to the cell only at the basal body.  The waves generated by the axonemal sliding can be either asymmetric or symmetric waves ~\cite{nate2}.  Recently, it was suggested that the central pair of microtubules rotate within the nine outer-doublet microbutules, activating their movements sequentially as the C1 microtubule of central pair is appropriately oriented ~\cite{nate3}.
	Classically, flagellar motility is associated with cell 'swimming' in a liquid media. However, recent experiments with Paranema indicate that flagella are also important for motile cells to move along solid supports ~\cite{nate4}.  Interestingly, similarly to the Trypanosomatids, the Paranema flagellum is attached to cell body along most of its length and contains a second filament within the flagellum called the paraflagellar rod (PFR).  Mutational analysis has shown the PFR is required for productive cell motility in African trypanosomes, American Trypanosomes (T. cruzi) and in Leishmania ~\cite{nate5}.  
With a variety of protozoan motility models choose from, we chose to utilize trypanosomes because of the specific advantages the system offers relative to the other model systems.   For example, although trypanosomes can not be used in classical genetic studies, the molecular biology tools (expression vectors, dsRNAi, etc) that are available for African trypanosomes far exceed those in any of the other protozoan models.  Additionally, the African trypanosome has an attached flagellum that allows us to observe wave propagations along the flagellar membrane.  The foundational cell biology of trypanosomes has been exquisitely characterized and offers other unique features (such the absence of any filamentous actin) that affect cell motility. 
The African trypanosome flagellum consists of a cononical 9+2 axoneme, the paraflagellar rod (PFR), and the flagellar adhesion zone (FAZ), all of which are required for successful propulsion ~\cite{nate6}.   The resulting movement of swimming trypanosome is an auger-like motility with the cell body following the free-end of the flagellum.  Recently, we observed that the movements of the flagellum are not simple wave oscillations, but rather appeared to be a complex variety of bends, twists, and waves.  In fact, we also noticed that trypanosomes exhibit a slithering-like motility in high viscosity liquid (unpublished observation), and a push rod like motility when covalently attached to a solid substrate (glass).  This is not surprising if light of the fact that  trypanosomes are capable of moving through virtually all non-mineralized biological tissues (including epithelium, muscle, and the blood-brain barrier).  

\section{Chemistry immobilization of Trypanosome}

The immobilization of whole, living cells is gaining importance because of two independent objectives.  One objective stems from the need to know the chemical and biological composition of our environment.  Indeed, with the recent scare from bio-weapons and the SARS virus, the detection of biological entities is gaining not only a scientific audience but a lay audience as well.  Biosensors are the subset of sensors, which take advantage of the exquisite specificity of biological entities such as antibodies and enzymes.  For example, antibodies are often immobilized onto sensing surfaces, because of their exquisite specificity-an antibody can distinguish a single protein from a population of 10 million proteins~\cite{darrell1}.  The immobilization schemes used to immobilize antibodies are varied:  Sepaniak and co-workers were the first to covalently immobilize antibodies onto a surface~\cite{darrell2}. while antibodies have been immobilized within a sol-gel matrix by Bright and co-workers~\cite{darrell3} and immobilized within polymer networks~\cite{darrell4}.  However, there are a very few research groups currently which are taking a different approach-that is to immobilize an entire cell onto a surface and to use the whole cell as a sensor.  Francis Ligler, at the Navel Research Lab has immobilized whole cells (but dead) for use with optical fiber biosensors in order to determine TNT concentration~\cite{darrell5}. The second objective stems from basic biological research.  For instance, Suzaki and co-workers immobilized living Peranema trichophorum using poly-L-lysine in order to study the gliding movement~\cite{darrell6}.   Furthermore, Engman and Tyler immobilized living Trypanosoma cruzi using a hydrophobic attraction in order to study how environmental change effects cell response~\cite{darrell6}.  A common thread exists between both biosensor development and fundamental biological research: the immobilization of living cells.

Here we report the two simple immobilization procedures to the immobilization of Trypanosome onto glass surfaces.  All chemicals were purchased from Sigma-Aldrich (St. Louis, MO.) and used without purification. The first chemistry is non-lethal and is a silane-amine-aldehyde-cell linkage.  Specifically, glass coverslips were incubated in chromic acid for 2 hours.  The slides were removed and dried.  Then approximately 250 microliters of 3-aminopropyltriethoxysilane were placed on the coverslips and allowed to sit for approximately 5 minutes.  The 3-aminopropyltriethoxysilane was removed via aspiration.  Next, approximately 250 microliters of gluteraldehyde (EM grade) was allowed to stand on the glass coverslips for 5 minutes.  Finally 200 microliters of cells at a density of $6.0 10^{5}$ cells/ml were allowed to incubate on the modified surface.  The net result is shown schematically in Figs. I and II.  The second immobilization chemistry is lethal and is a silane-cell linkage.  Glass coverslips were incubated in chromic acid for 2 hours.  The slides were removed and dried.  Then approximately 250 microliters of 3-aminopropyltriethoxysilane were placed on the coverslips and allowed to sit for approximately 5 minutes.  Finally 200 microliters of cells at a density of $6 10^{5}$cells/ml were allowed to incubate on the modified surface.

\section{Physical model}

Flagella are enlongated elastic systems which can generate motion and self-propulsion in the surrounding fluid.  The axoneme is characterized by nine parallel pairs of microtubules arranged in a circular fashion together with a large number of dynein molecular motors ~\cite{andy1}. In this work we introduce a two-dimensional physical model for flagellum dynamics. Our approach is inspired by the theory of elastic deformation of beams with continuous distribution of internal forces and torque produced by the molecular motors ~\cite{andy2}. We consider two parallel straight lines of length $L$ separated by a distance $a$. When we introduce the dynamics, these two lines are deformed into two parallel arbitrary curves (same separation distance $a=a(s)$) ${\vec r}(s)$ and ${\vec r}(s)-a(s){\vec n}(s)$, parametrized by the natural arc length parameter $s$. Our definition for "parallel" curves is given below. In this deformation we consider that at the initial point ($s=0$)the shift between curves is always zero. From differential geometry we know that the local shift $\delta (s)$ that occurs between the parallel curves is given by the differential equation
$$
{{d\delta}\over{ds}}={{a(s)}\over {R(s)}},
$$
where ${1 \over {R(s)}}=|{{d^{2}{\vec r}}\over {ds^2 }}|$ is the local curvature, and ${\vec n}(s)=R(s){{d^{2}{\vec r}}\over {ds^2 }}$ is the normal to the ${\vec r}(s)$ curve. Consequently, in any point of the pair of parallel curves we have a total shift given by
\begin{equation}
\delta (s) =\int_{0}^{L}{{a(s)}\over{R(s)}}ds.
\label{eq:shift}
\end{equation}

If we consider elastic ends (free end at $s=L$) for the pair of strings (the curves described above) there is no restriction for the shape. If one imposes rigid ends, one has to use only zero total curvature shapes, unless the shift produced by curvature is compensated by a variable distance between the curves. In other words, a constant diameter flagellum with rigid ends cannot have any shape. In the following we consider an uniform distribution of molecular motors, between the two lines (initial moment). When the motors ar activated they start to crawl along the lines and to extend their lengths. This dragging force is balanced by the elastic contribution of the strings and consequently the system starts  to have a non-zero curvature. In Fig. 4 we present such a scenario for the deformation of the lines into an arch of circle. Each arrow describes actually several motors, so we have variable magnitude for the local displacement. In the end, because of the constant curvature on the circle, the spacing is again equal.

From general theory of elastic beams ~\cite{andy3} we have the distribution of infinitezimal torque along the beam given by
\begin{equation}
{\vec M} =\epsilon _{lij}\int_{\partial V}(x_i \sigma_{jk}-x_j \sigma_{ik}) dS_{k}{\vec e_{l}},
\label{eq:torque}
\end{equation}
where ${\vec e}_i$ are the versors of the coordinate system, $dS$ is the area element, $V$ is the volume of the beam, and $\sigma _{ij}$ is the stress tensor. For thin beams the local deformation ${\vec u}({\vec r})$ in given by 
$$
u_x =-{{1}\over {2R}}(z^2+\sigma (x^2-y^2)),
$$
\begin{equation}
u_y =-\sigma {{xy}\over{R}}, \ \ \ \ \ u_z =\sigma {{xz}\over{R}},
\label{eq:strain}
\end{equation}
where $\sigma$ is the average value of the diagonal elements of the stress tensor for a homogenous material. By introducing Eq.~(\ref{eq:strain}) in Eq.~(\ref{eq:torque})
we obtain a simpler equation for the local torque in a 2-dimensional motion, namely
\begin{equation}
M_y = {E \over R} \int_{V} x^2 dS={{E J_{y}}\over R},
\label{eq:strain2}
\end{equation}
where $E$ is Young modulus of elasticity, $J_y$ is the moment of inertia of the beam, and we have chosen deformation in the $x-z$-plane.

The dynamical equation for the local transversal deformation $u(s)$ of a beam under the action of force field ${\vec F}$, in the limit of small deformations is described by 
\begin{equation}
{{\partial ^2 }\over{\partial s^2}}(E J_y {{\partial ^2 u}\over {\partial s^2}})- 
{{\partial }\over{\partial s}}(P {{\partial  u}\over {\partial s}})+m
{{\partial ^2 u }\over{\partial t^2}}=F,
\label{eq:dynlin}
\end{equation}
where we have also included additional axial pressure $P$. Here $m$ is the mass per unit of length. In the approximation of constant parameters along the beam, that is $E$ and $J_y$ we can write the equation for general deformations. In this case a nonlinear term is introduced by the curvature
\begin{equation}
{{\partial ^4 u}\over{\partial s^4}}- {{P}\over {E J_y}} \biggl (
{{\partial ^2 u }\over{\partial s^2}}-{3 \over 2}\biggl ( {{\partial u}\over {\partial s}}\biggr ) ^2{{\partial ^2 u}\over {\partial s^2}}\biggr ) +{m \over {E J_y}}
{{\partial ^2 u }\over{\partial t^2}}={F\over {E J_y}}.
\label{eq:dyn}
\end{equation}
Eq.~(\ref{eq:dyn}) is nonlinear partial differential equation which allows localized solutions in the form of solitary waves. For example if we choose a soliton-like shape for initial data ~\cite{andy4}, the numerical solution shows a conservation of the localization for long time scale, enough to allow the perturbation to travel along the flagellum with a speed of the order of surface linear waves. 

In the following we have investigated the displacement and rotation of a system cell+flagellum during a relative rigid rotation of the flagellum. By taking into account the conservation of center of mass and angular momentum, we can make an estimation of the change of direction of the cell. We approximated the cell with an ellipsoid of semiaxes $A,B,B$ and density $\rho _{C}$, and the flagellum by a rigid rode of length $L$, diameter $a$, and density $\rho_{F}$. For a relative rotation angle $\alpha$ between the long ellipsoid axis and flagellum we have a translation of the system ${\vec {\tau}}$ and an absolute  rotation of the cell $\psi $. The masses and moments of inertia of the cell and flagellum are approximated with 
\begin{equation}
m_C={{4 \pi A B^2}\over 3}  \rho_{C}, \ \ \ m_F=\pi a^2 L \rho_F, \ \ \ \  J_C={{(6a^2+B^2 m_C} \over {5}}, \ \ \  J_F={{L^2 m_F}\over {3}}.
\label{eq:moment}
\end{equation}
The global translation of the system is given by
$$
\tau_x={{A cos {{\beta \alpha}\over{\beta -1}}-{1 \over 2}\epsilon L cos{{\alpha}\over{\beta - 1}}}\over {1+\epsilon}}
$$
\begin{equation}
\tau_y={{A \biggl (sin {{\beta \alpha}\over{\beta -1}} -1 \biggr ) +{1 \over 2}\epsilon L\biggl (1- sin {{\alpha}\over{\beta - 1}}\biggr ) }\over {1+\epsilon}},
\label{eq:tra1}
\end{equation}
where $\beta=J_F / J_C $, $\epsilon =m_F /m_C$. The absolute rotation angle of the symmetry axis of the cell is given by
\begin{equation}
\psi ={{\beta  (\alpha +(\beta -1){{\pi}\over {2})}}\over {\beta -1}} -(\beta -1){{\pi}\over{2}}.
\label{eq:rota}
\end{equation}
From these equations one can make average estimation for the speed and the angle of turning of the cell during a "rigid rotation" type of motion. In these calculations the viscosity of the fluid around the system was neglected. We use the following values for the parameters: $A=20\mu m$, $B=5\mu m$, $L=3\mu m$, $a=0.7\mu m$, $\rho_{C}=478$ $kDa/\mu^3$=$0.793$ $g/cm^3$. By correlating these values with the kinematic data obtained from experiment (Fig. III), namely angular speed of rotation of about $\omega = 8.7 - 9.2$ rad/s, we obtain a value for the density of the flagellum of $\rho _F = 1,020$ $ kDa/\mu^3$ =$2.14$ $g/cm^3$.
A picture of a rotation of $\alpha =\pm 30^o$  is presented in Fig. 3.

In Fig.4 we present specific values of translation and rotation of the cell+flagellum system for relative flagellar rotation between $-45^o$ and $45^o$. The correlation between rotation and translation is in very good agreement with experimental data.

\section{Results}

By examining the movements of individual procyclic form Trypanosoma brucei brucei (strain Ytat1.1), we were able to characterize five distinct modes of flagellar motility.  Three of the five modes are bending-like movements and two of the modes are wave-like undulations (Fig. III).  In this figure the left column graphically describes the general mode of the flagellum.  The right columns are sequential images of a time-lapse series (0.125 sec/frame). DIC time-lapse images were captured on an Olympus BX60 microscope equipped with 'The Spot'™ CCD.  Individual images were processed in Adobe Photoshop 5.5 (Palo Alto, CA).

The bending-like movements are described as 'elastic vibration', 'rigid translation', and 'rotation'.  Rigid oscillation of the flagellum is characterized by a primarily two-dimensional whip-like movement of the distal end of the flagellum relative to the rest of the tethered cell (Fig. IIIA).  During elastic vibrations, the flagellum is flexible, as indicated by changes in shape along the length of the flagellum, but the motile force appears to be generated by a pivot or oscillating origin near the mid-point of the cell. The dashed line in 'A' show the original position of the flagellum, and the arrow head in indicates the tip of the flagellum in each frame.  Likewise, during 'rigid translation' movements of the flagellum, the movement of the flagellum is restricted to bending at the mid-point of the cell, but in this case the flagellum does not change shape along its length during (Fig. IIIB). The perpendicular in 'B' provides a point of reference for the bending flagellum, and the arrow indicates the original position of the flagellar tip in the first frame. The rotational mode of flagellar motility is characterized by a 'rigid translation'-like motion with an angular displacement that causes the path of the flagellum to be conical.  The dashed circle in 'C' indicates the rotational path of the flagellar tip, and the arrow indicates the direction of rotation.  The modes of motility described in A, B, and C do not appear to contain wavelike undulations.  This type of motility resembles that of the whip-like rotation of a bacterial flagellum (Fig. IIIC).  The rotation mode is most often observed when there is a bend in the flagellum between the distal tip of the flagellum and the mid-point of the cell. All three bending modes can persist for several seconds without breaking the symmetry. The 4th type of motion, namelly  'wave propagation' type flagellar motility is characterized by symmetric or asymmetric undulations of the flagellum. The arrows in each frame indicate the phase of the wave.  We have identified a 4th possible type of motion, the 'depression wave' movements which are characterized by a single membrane depression moving along the flagellum without a symmetric oscillating wave following it.  Arrows indicate the depression in the membrane, and the brackets indicate the movement of depression (notice the lack of another depression behind the arrow). The cell in 'E' (the 5th type of motion) is had nearly completed cell division.  The depression waves are typically observed only briefly between transitional wave patterns or changes in modes of motility. 
The last two modes of motility that we have characterized in this study are the 'wave propagation' and 'depression wave' ~\cite{andy5}. The 'wave propagation' type flagellar motility is characterized by symmetric or asymmetric undulations of the flagellar membrane.  This mode of motility is the typical sinus wave-like motion that is classical associated with the eukaryotic flagellum.  Interestingly, the amplitude, period, and frequency of the waves can be highly variable, even within sequential frames of the same time series.  Wave propagation can persist for relatively long periods, although variations in frequency are often noticeable.  During the dynamic periods of wave propagation, we noticed the 'depression wave' mode, which is characterized by a single membrane depression moving along the flagellum without an oscillating wave following it. This type of wave is observed only briefly (but consistently) between transitional wave patterns or different modes of motility. 
When monitoring an individual cell for extended periods (minutes - hours), we can observe many different modes motility that occur at different frequencies and persist for different amounts of time.  We are currently quantifying the relative ratio of each mode of motility in individual cells and within populations of cells.

From the point of view of the dynamical model (physical model) we note that for a given shape of the flagellum we can provide the average distribution of active molecular motors inside it. For exampl, and experimental snap shot of the motion can be interpolated with analytic functions in order to obtain the mathematical form of the curve. For this curve we can calculate the local curvature which is responsible for the action of molecular motors. Both the local torque produced by the motors, and the relative shift are proportional with the curvature. In Fig.2 we present such examples.

\section{Experiment-model comparison and predictions}

We measured in 13 movies (in about 2000 frames) regular patterns in about 400 frames, that is $20\%$  of the total movie volume. The relative error was about $10\%$, but this can be improved by a better statistics. The amplitude of motions ranges from A=0.5-1.2m for waves and oscillations to 7-10m for rotation, helix and translation motions. In all movies we detected in a reproducible way wavelength in the range of 4-5m  for traveling waves, to 7m for stationary waves, both for waves and breaking-symmetry modes. The frequency of periodic motions ranges from 1.33 - 1.4 Hz for waves and rotations, to 4 HZ for rigid moves. The speed has a large range. For wave-like, elastic oscillations, and rigid translation motions we have in general 2-8 m/s. For helix-like rotations the speed is larger, like v=10-11 m/s, and for free almost moving cells we measured about 60 m/s. We also mention the possibility of having nonlinear behavior through breaking of symmetry between different modes (see also ~\cite{andy2}). We noticed an interesting breaking of a 2 mode in a 4 mode, that is a frequency doubling effect which occurs only for nonlinear dynamical systems.  

The most important phenomenon predicted by our theory, and put into evidence experimentally is the occurance of a depression wave along the cell-attached part of the flagellum (see Fig. III). The dynamics of the system can be modeled by Eq. (6) which is a nonlinear partial differential equation in time-space for the deformation of the flagellum with a fixed end. If denote ${{\partial u}\over{\partial x}}=\eta (x,t)$, and we look for traveling solutions of the form $\eta (x,t)=f (x-V t)$, where $V$ is the supposed velocity of the wave, Eq.(6) reads:
\begin{equation}
f_{\xi \xi \xi}-{P\over {E J}}(f_{\xi \xi}- {3 \over 2}f^2 f_\xi )+ {{m}\over {E J}}={F\over {E J}}
\label{eq:asol}
\end{equation}
where $\xi =x-Vt$. This equation is nothing but the most general, and at the same time the simplest, version of a 1-dimensional nonlinear model, namely a Korteweg-de Vries equation (terms 1,3, and 4) with dissipation (term 2) and external sources (RHS). If we neglect the non-homogenous term provided by the internal forces of molecular motors (relaxed motors) we have a dissipative KdV~\cite{andy4,andy5}. The critical dissipation term can be tuned to have either positive or negative value by tuning the elongation force $P$ acting at the cell-attached end of the flagellum. Consequently we can have either solitons, or anti-solitons (i.e. the observed depression waves) along the flagellum, function of the pressure/depression boundary condition produced by the interaction with the cell. Of course, there is still much to be checked, especially since in that investigated part, the flagellum is fully attached to the membrane, but an intuitive picture about such localized traveling waves (which can also bring some hint concerning the coherence and synchronization of molecular motor action) can be proved.

\section{Conclusion}

In this paper we describe how chemical immobilization facilitates investigating flagellar motion, and we present a physical model for these types of motion  We found at least 5 distinct modes of flagellar motility in procyclic form African trypanosome and we note that Trypanosoma brucei flagellar beating is a complex combination of different modes of beating/bending.The physical model includes a nonlinear dynamical equation which can justify the occurence of stable, localized traveling depression waves, among other linear waves. This work is just the begining of a series of papers in which we will describe in detail the nonlinear dynamics and approaches to a full three-dimensional one. In the end we want to address several open questions raised by this work.  Are the different modes of flagellar beating are differentially regulated by temporal, spatial, and/or chemical signals. Does the axoneme beat at a constant frequency?Are different trypanosome flagellar mutants defective in different modes of motility? Do chemically immobilized cells behave different than electrostatically.?Do motility-altering chemicals affect different modes of motility? In addition we would like to develope in the future the following issues: Flagellar mutants, the role of the PFR in motility, description of tension and slack within the flagellar membrane, and limitations of cell 'position' and tethering, angle of view. Forthcoming papers will try to give partial answer to these questions.

\section*{Acknowledgments}
The contribution work of one of us (AL) was supported by the National Science Foundation under NSF Grant 0140274. We greatly appreciate the work of our undergraduate research assistants A. Jones, L. Archuleta, J. Rains, J. Cain, A. Dunham, H. Moffett, and C. Radcliff.

\newpage

{\bf Figure Captions}

\vspace{2cm}

Fig. I and II

Schematic result of chemical immobilization of cells.

\vspace{0.5cm}

Fig. III

Characterizing five different modes of flagellar motility in procyclic African trypanosomes.  The left column graphically describes the general mode of the flagellum.  The right columns are sequential images of a time-lapse series (0.125 sec/frame).  A. 'Elastic vibration' of the flagellum is characterized by a whip-like movement of the distal end of the flagellum relative to the rest of the tethered cell.  The dashed line in 'A' show the original position of the flagellum, and the arrow head in indicates the tip of the flagellum in each frame. B. 'Rigid Translation' of the flagellum is characterized by a stiff flagellum that is moving due to flagellar bending near the mid-point of the cell body. The perpendicular in 'B' provides a point of reference for the bending flagellum, and the arrow indicates the original position of the flagellar tip in the first frame. C. 'Rotation' of the flagellum is characterized by a 'rigid translation'-like motion with an angular displacement that causes the path of the flagellum to be conical. The dashed circle in 'C' indicates the rotational path of the flagellar tip, and the arrow indicates the direction of rotation.  The modes of motility described in A, B, and C do not appear to contain wavelike undulations.  D. 'Wave propagation' type flagellar motility is characterized by symmetric or asymmetric undulations of the flagellum. The arrows in each frame indicate the phase of the wave.  E. 'Depression wave' movements are characterized by a single membrane depression moving along the flagellum without a symmetric oscillating wave following it.  Arrows indicate the depression in the membrane, and the brackets indicate the movement of depression (notice the lack of another depression behind the arrow). The cell in 'E' is had nearly completed cell division.  The depression waves are typically observed only briefly between transitional wave patterns or changes in modes of motility. DIC time-lapse images were captured on an Olympus BX60 microscope equipped with 'The Spot'™ CCD.  Individual images were processed in Adobe Photoshop 5.5 (Palo Alto, CA).

\vspace{0.5cm}

Fig. 1

The deformation and relative shift of two parallel filaments from straight line into circular pattern. The arrows represent the action of molecular motors. The local relative shifts are equal in the end because of the constant curvature, while in between the shifts have variable magnitudes. 

\vspace{0.5cm}

Fig. 2

We present the distribution of molecular motor torque along the filaments for three examples of filament patterns: Damped sinus oscialltion, localized deformation with zero global curvature, and localized deformation with non-zero global curvature. 

\vspace{0.5cm}

Fig. 3

Displacement and rotation of the system cell+flagellum for two symmetric relative rotations of the flagellum of $60^o$ each. Conservation of momentum and angular momentum results in possible change of direction of the swimming. 

\vspace{0.5cm}

Fig. 4

Plot of displacement of center of mass and rotation of the system cell+flagellum versus relative rotation of the flagellum with respect to the cell axis of symmetry.


\begin{thebibliography}{0}

\bibitem{nate1}
Dutcher SK. ,Trends Genet. 1995 Oct;11(10):398-404. Review.

\bibitem{nate2}
Wakabayashi K, Yagi T, Kamiya R., Cell Motil Cytoskeleton. 1997;38(1):22-8.

\bibitem{nate3}
Wargo MJ, Smith EF., Proc Natl Acad Sci U S A. 2003 Jan 7;100(1):137-42. Epub 2002 Dec 23.

\bibitem{nate4}
Saito A, Suetomo Y, Arikawa M, Omura G, Khan SM, Kakuta S, Suzaki E, Kataoka K, Suzaki T. . Cell Motil Cytoskeleton. 2003 Aug;55(4):244-53.

\bibitem{nate5}
Maga JA, LeBowitz JH.,  Trends Cell Biol. 1999 Oct;9(10):409-13. Review; Bastin P, Pullen TJ, Sherwin T, Gull K., J Cell Sci. 1999 Nov;112 ( Pt 21):3769-77; Bastin P, MacRae TH, Francis SB, Matthews KR, Gull K. , Mol Cell Biol. 1999 Dec;19(12):8191-200.

\bibitem{nate6}
Bastin, P., Z. Bagherzadeh, et al. (1996). ,  Mol Biochem Parasitol 77(2): 235-9; Gull, K. (1999)., Annu Rev Microbiol 53: 629-55; Hutchings, N. R., J. E. Donelson, et al. (2002). , J Cell Biol 156(5): 867-77;  Kohl, L. and K. Gull (1998).,  Mol Biochem Parasitol 93(1): 1-9; Kohl, L., T. Sherwin, et al. (1999)., J Eukaryot Microbiol 46(2): 105-9.

\bibitem{darrell1} Benjamine, E. and S. Leskowitz, Immunology: A hort Course.  2nd edition.  1991, New York: Wiley-Liss.  Pg 459.

\bibitem{darrell2} Tromberg, B.J., et al.  Analytical Chemistry, 1987  59 (8): p 1226-1230; Vo-Dinh, T., et al.,   Applied Spectrscoy, 1987 41(51):p. 735-738.

\bibitem{darrell3} Ingersoll, C.M. and F.V. Bright,  Chemtech, 1997 27(1): p. 26-31; MacCraith, B.J.,  Sensors and actuators B, 1993.  11: p.29-34.

\bibitem{darrell4} Disley, D.M., et al., Biosensors and Bioelectronics, 1998 13(3-4): p. 383-396.

\bibitem{darrell5} L.C. Shriver-Lake, K.A. Breslin, J.P. Golden, F. S.  Ligler.  (2001), "Fiber optic biosensor for the detection of TNT" pp.52-58 SPIE Proceedings: Vol. 2367:52-58.

\bibitem{darrell6} Saito, Akira et al., Cell Motility and the Cytoskeleton 55:244-253(2003).  

\bibitem{darrell7} Tyler, Kevin and Engman, David, Cell Motility and the Cytoskeleton 46:269-278(2000).  

\bibitem{andy1}F. J\"ulicher, and J. Prost, {\em Phys. Rev. Lett.} {\bf 75}, 2618 (1995) and  {\bf 78}, 4510 (1997)

\bibitem{andy2}S. Camalet, F. Julicher, and J. Prost, {\em Phys. Rev. Lett.} {\bf 82}, 1590 (1999); L. Bourdieu, {\em et al}, {\em Phys. Rev. Lett.} {\bf 75}, 176 (1995); K. Sekimoto, N. Mori, K, Tawada, and Y. Y. Toyoshima, {\em Phys. Rev. Lett.} {\bf 75}, 172 (1995)

\bibitem{andy3}L. Landau and E. Lifchitz, {\em Theorie de Elasticite} (MIR, Moscow, 1967); E. B. Magrab, {\em Vibrations of Elastic Structural Members} (Sijthoff and Noordhoff, Germantown, MD 1979); E. Skudrzyk {\em Simple and Complex Vibratory Systems} (The Pennsylvania State Univ. Press, University Park, 1968)

\bibitem{andy4}D. J. Korteweg and G. de Vries, {\em Phil. Mag.} {\bf 39}, 422L (1895); M. Remoissenet, {\em Waves Called Solitons} (Springer-Verlag, Berlin, 1999)

\bibitem{andy5} A. Ludu, and J. P. Draayer, {\em Phys. Rev. Lett.} {\bf 80}, 2125 (1998)

\end{thebibliography}
\end{document}